\documentclass[preprint,3p,twocolumn]{elsarticle}
\usepackage{geometry}
\usepackage{fancyhdr}

\biboptions{numbers,sort&compress}
\usepackage{graphicx}
\usepackage{amsmath}
\usepackage{bm}
\usepackage{url}
\begin{document}

\begin{frontmatter}

\title{Three-dimensional needle network model for dendritic growth with fluid flow}

\author{ Thomas Isensee$^{\,1,2}$, Damien Tourret$^{\,1}$}
\address{$^{1}$\,IMDEA Materials Institute, Getafe, 28906 Madrid, Spain}
\address{$^{2}$\,Polytechnic University of Madrid, 28040 Madrid, Spain}
\address{\rm\url{damien.tourret@imdea.org}}

%%%%%%%%%%%%%%%%%%%%%%%%%%%%%%%%%%%%%%%%%%%%%%%%%%%
\begin{abstract}

We present a first implementation of the Dendritic Needle Network (DNN) model for dendritic crystal growth in three dimensions including convective transport in the melt. The numerical solving of the Navier-Stokes equations is performed with finite differences and is validated by comparison with a classical benchmark in fluid mechanics for unsteady flow. We compute the growth behavior of a single equiaxed crystal under a forced convective flow. As expected, the resulting dendrite morphology differs strongly from the case of the purely diffusive regime and from similar two-dimensional simulations. The resulting computationally efficient simulations open the way to studying mechanisms of microstructure selection in presence of fluid flow, using realistic alloys and process parameters. 
\end{abstract}

\end{frontmatter}
\thispagestyle{fancy}

%%%%%%%%%%%%%%%%%%%%%%%%%%%%%%%%%%%%%%%%%%%%%%%%%%%
\section{Introduction}
\label{sec:Intro}

Dendritic microstructures are common in solidification-processed metals and alloys \cite{langer1980,trivedi1994}. Their morphology has crucial influence on the thermo-mechanical properties of these materials \cite{dantzig2009}. Dendritic patterns are shaped by the interaction of individual dendritic branches and hence arise from mechanisms on different length scales, i.e. phenomena on the scale of the microscopic solid-liquid interface, and macroscopic heat and solute transport. 
For decades, the aim to bridge scales has motivated the development of a wide range of multiscale simulations approaches, e.g. using dedicated formulations for enthalpy terms~\cite{voller1987fixed}, volume-averaged conservation equations~\cite{wang1996equiaxed}, cellular automata coupled with finite elements~\cite{gandin1999three} or finite differences~\cite{wang2003model}, or dynamics of grain envelopes~\cite{steinbach1999three}, to name a few. 
Macroscopic heat and solute transport can  be strongly influenced by advective currents. Since buoyancy can be caused by gravity \cite{mehrabian1970,nguyen1989,dupouy1989}, it is almost impossible to carry out solidification experiments in homogeneous conditions. Experiments with reduced gravity have provided important insight into the effect of buoyancy onto microstructure selection \cite{glicksmann1994,nguyen2005,nguyen2017}, however only at great expense, which is why the modeling of dendritic growth under fluid flow is of tremendous interest.

In contrast to growth in the purely diffusive regime, in the presence of fluid flow, growing crystals are asymmetric. The problem has been investigated analytically, e.g. using a diffusive boundary layer (``stagnant film'') around the dendrite \cite{cantor1977,li2002,sekerka1995}, and computationally, e.g. with models that explicitly track the solid-liquid surface \cite{udaykumar2003,alrawahi2002,zhao2005} or phase-field (PF) simulations \cite{beckermann1999,tong2001,jeong2001,jeong2003,lu2005,rojas2015}.
However, these models still require a reasonably accurate resolution of the dendrite tip morphology, which limits their use in the case of concentrated alloys that usually solidify with thin needle-like branches with a tip radius much lower than the scale of macroscopic transport of heat and solute in the melt.

To address this multiscale problem we developed a {\it dendritic needle network} (DNN) model, in which the dendrite branches are represented as thin paraboloids. The model quantitatively predicts the dynamics of individual branches in complex hierarchical networks during alloy solidification at a scale much larger than the diffusion length, and it is particularly adapted to solidification at low supersaturation, i.e. low P\'eclet number, when phase field simulations become prohibitively costly. 
The key computational advantage of the method comes from the fact that the numerical grid spacing can be chosen of the same order as the dendrite tip radius, which is typically one order of magnitude coarser than the requirement for quantitative phase-field simulations. This results in simulations faster by several orders of magnitude, with minimal loss in accuracy~\cite{tourret2020}. 
The DNN model, developed in two dimensions (2D) \cite{tourret2013a} and three dimensions (3D) \cite{tourret2016}, was already verified against classical theories and phase-field simulations of dendritic growth, and validated against directional solidification experiments in a predominantly diffusive transport regime \cite{tourret2016,tourret2015,tourret2015jom}. A first incorporation of fluid flow in 2D was also achieved \cite{tourret2019}. However, since flow patterns strongly differ in 2D and 3D, quantitative simulations require a 3D implementation. 
Here we present first results of the DNN model including fluid flow in 3D.

%%%%%%%%%%%%%%%%%%%%%%%%%%%%%%%%%%%%%%%%%%%%%%%%%
\section{Model}
\label{sec:model}

\subsection{Sharp-interface problem}

Considering a binary alloy at a temperature $T=T_0$ below its liquidus temperature $T_\text{L}$, we introduce a dimensionless form of the solute concentration field $c$, i.e. the supersaturation

\begin{equation}
 U := \frac{c_0 - c}{(1-k)c_0}
\end{equation}
with $c_0$ and $k$ the liquid equilibrium concentration at $T=T_0$ and the interface solute partition coefficient, respectively. Neglecting diffusion in the solid phase, the solute concentration in the liquid close to the solid-liquid interface (e.g. within a diffusive boundary layer) follows 
\begin{equation}
\label{eq:diffusion}
 \partial_t U = D\nabla^2 U.
\end{equation}
Solute mass conservation at the interface takes the form of a Stefan condition
\begin{equation}
\label{eq:stefan}
 \bm{v}_n = D\partial_nU|_i\,\text{,}
\end{equation}
that relates the interface normal velocity $\bm{v}_n$ to the normal solute concentration gradient $\partial_n U|_i$. Neglecting kinetic undercooling, the Gibbs-Thomson relation for interfacial equilibrium is
\begin{equation}
\label{eq:equilibrium}
 U_i = d_0 f_\gamma(\overline{\theta})\kappa,
\end{equation}
where $\kappa$ is the interface curvature, $d_0 = \Gamma_{sl}/\left[|m|(1-k)c_0\right]$ the solutal capillary length with the liquidus slope $m<0$ and the Gibbs-Thomson coefficient $\Gamma_{sl}$. The anisotropy function $f_\gamma(\overline{\theta})$ represents the dependence of the interface stiffness $\gamma(\overline{\theta}) + \partial^2_{\theta}\gamma(\overline{\theta}) = \gamma_0 f_\gamma(\overline{\theta})$ upon the interface orientation $\overline{\theta}$, where $\gamma(\overline{\theta})$ is the excess free-energy of the solid-liquid interface and $\gamma_0$ is its average value \cite{haxhimali2006,dantzig2013}. Eqs. \eqref{eq:diffusion} - \eqref{eq:equilibrium} describe the sharp interface problem for a growing solid-liquid interface, combined with an imposed supersaturation
\begin{equation}
 \Omega := \frac{c_0 - c_\infty}{(1-k)c_0}
\end{equation}
far away from the interface, with $c_\infty$ the nominal solute concentration.

\subsection{DNN model}

The DNN model is designed to model solidification at low supersaturation, i.e. at a P\'eclet number $\text{Pe} = RV/(2D)\ll1$, with $R$ and $V$ the dendrite tip radius and velocity, respectively, and $D$ the liquid solutal diffusion coefficient. 
With a tip radius much smaller than the diffusion length $l_D = D/V$, conservation equations can be derived at different length scales, such that: (i) mass conservation on an intermediate scale between the tip radius $R$ and the diffusion length, provide the product $RV$ (or $RV^2$ in 2D), and (ii) microscopic solvability theory at the scale of $R$ prescribes the constant value of $R^2V$. The combination of those conditions, described below, enables to obtain the instantaneous growth conditions $R(t)$ and $V(t)$ for each dendritic tip.

\subsection{Microscopic solvability condition at the scale of the dendrite tip}

The well-established solvability theory \cite{langer1987,langer1989,barbieri1989,benamar1993} states that the sharp-interface problem described by eqs. \eqref{eq:diffusion} - \eqref{eq:equilibrium} has a steady solution only if
\begin{equation}
\label{eq:microscopic_solvability}
 R^2V = \frac{2Dd_0}{\sigma},
\end{equation}
where the selection parameter $\sigma$ is uniquely determined by the strength of the interface crystalline anisotropy. The theory was confirmed using PF simulations \cite{karma1998,provatas1998,plapp2000a}, which even showed that the product $R^2V$ relaxes to a constant very early during growth, while $R(t)$ and $V(t)$ still undergo significant variations \cite{plapp2000a}. 
Analytical \cite{bouissou1989} and phase-field \cite{tong2001,jeong2003,lu2005} studies have also shown, that $\sigma$ remains constant up to fluid velocities about one order of magnitude higher than the tip velocity.
Thus, the DNN model considers eq. \eqref{eq:microscopic_solvability} to be valid at all times.

\subsection{Conservation of solute on an intermediate scale}

\begin{figure}[t]
\centering
\includegraphics[width=1.75in]{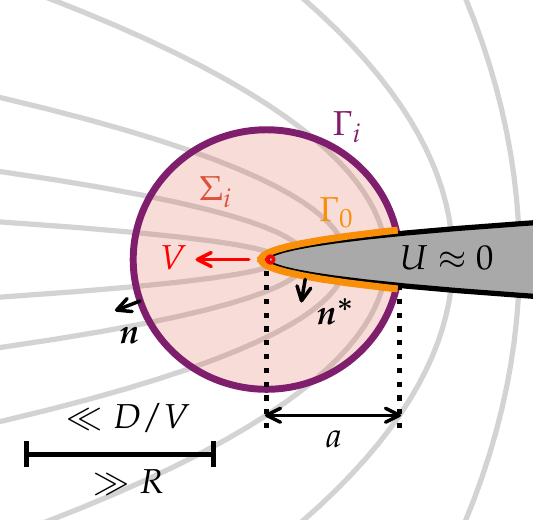}
\caption{Conservation of solute on the needle tip moving with velocity $V$ on an intermediate scale larger than the tip radius $R$ but smaller than the diffusion length $l_D = D/V$.
}
\label{fig:fif_0}
\end{figure}

On a scale larger than the tip radius, dendritic branches appear sharp and curvature effects can be neglected, so that the equilibrium concentration $U = U_i \approx 0$ is assumed along the interface. Assuming a paraboloid with its tip at $x_t$ and a cross section of radius $r_i(x) = \sqrt{(2R(x_t-x))}$ growing in a shape-preserving manner with velocity $V$, the mass conservation (eq. \eqref{eq:stefan}) integrated over a contour $\Gamma_0$ that spans until a length $a$ behind the tip (see figure \ref{fig:fif_0}) yields
\begin{align}
 D\int_{\Gamma_0} (\partial_n U) d\Gamma_0 &= \int_0^{2\pi}\int_0^{r_a} V r\,dr\,d\theta = 2\pi a R V ,
\end{align}
where $r_a$ is the cross section of the paraboloid at $x = x_t - a$. By introducing the flux intensity factor (FIF)
\begin{equation}
 \mathcal{F} := \frac{1}{2\pi a}\iint_{\Gamma_0} (\partial_n U)\,d\Gamma_0,
\end{equation}
\noindent the product $RV$ can be expressed as
\begin{equation}
RV=D\mathcal{F} .
\end{equation}
To integrate over $\Gamma_0$, we use the divergence theorem and the assumption of a Laplacian field in a moving frame with velocity $V$, i.e. $D\nabla^2U=-V\partial_xU$, yielding
\begin{equation}
 \iint_{\Gamma_0} (\partial_{n^*}U)\, dS = \iint_{\Gamma_i} (\partial_n U)\, dS +\frac{V}{D}\iiint_{\Sigma_i}(\partial_x U)\,dV.
 \end{equation}

\subsection{Navier-Stokes equation in the liquid phase}

We consider the liquid phase to be an incompressible and newtonian fluid, in which the mass conservation results in the incompressibility condition $\nabla\cdot\bm{u} = 0$. The conservative Navier-Stokes equations as a statement of momentum conservation read
\begin{equation}
 \rho\left[\partial_t\bm{u} + \nabla\cdot(\bm{u}\bm{u})\right] = \bm{F} - \nabla p + \eta\nabla^2\bm{u},
\end{equation}
where $\bm{u}$ is the velocity field of the fluid, $p$ is the pressure field, $\eta$ the dynamic viscosity, $\rho$ the fluid density and $\bm{F}$ represents external body forces, e.g. gravity-induced buoyancy.
The transport of solute within the liquid is described by the advection-diffusion equation

\begin{equation}
 \partial_t U + \nabla\cdot(\bm{u}U) = \nabla(D\nabla U).
\end{equation}

\subsection{Implementation}

Equations are solved using a method previously presented in 2D \cite{tourret2019}.
We use a projection method to solve of the momentum equation, with an iterative successive over relaxation method \cite{tourret2019} for the incompressibility condition $\nabla\cdot\bm{u} = 0$.
Space is discretized using finite differences on a homogeneous grid of cubic elements with staggered velocity components.
Time stepping uses an explicit Euler method.
The flux intensity factor $\mathcal{F}$ is integrated over a sphere centered around the tip location, similarly as done with a circle in 2D \cite{tourret2019}.
The code is implemented in C-based cuda programming language, which allows for parallelization on Nvidia graphic cards.

%%%%%%%%%%%%%%%%%%%%%%%%%%%%%%%%%%%%%%%%%%%%%%%%%

\section{Code validation}
\label{sec:validation}

\subsection{Usteady flow past an obstacle}
\label{subsec:karman}

\begin{figure*}[t]
\begin{centering}
\includegraphics[width=.7\textwidth]{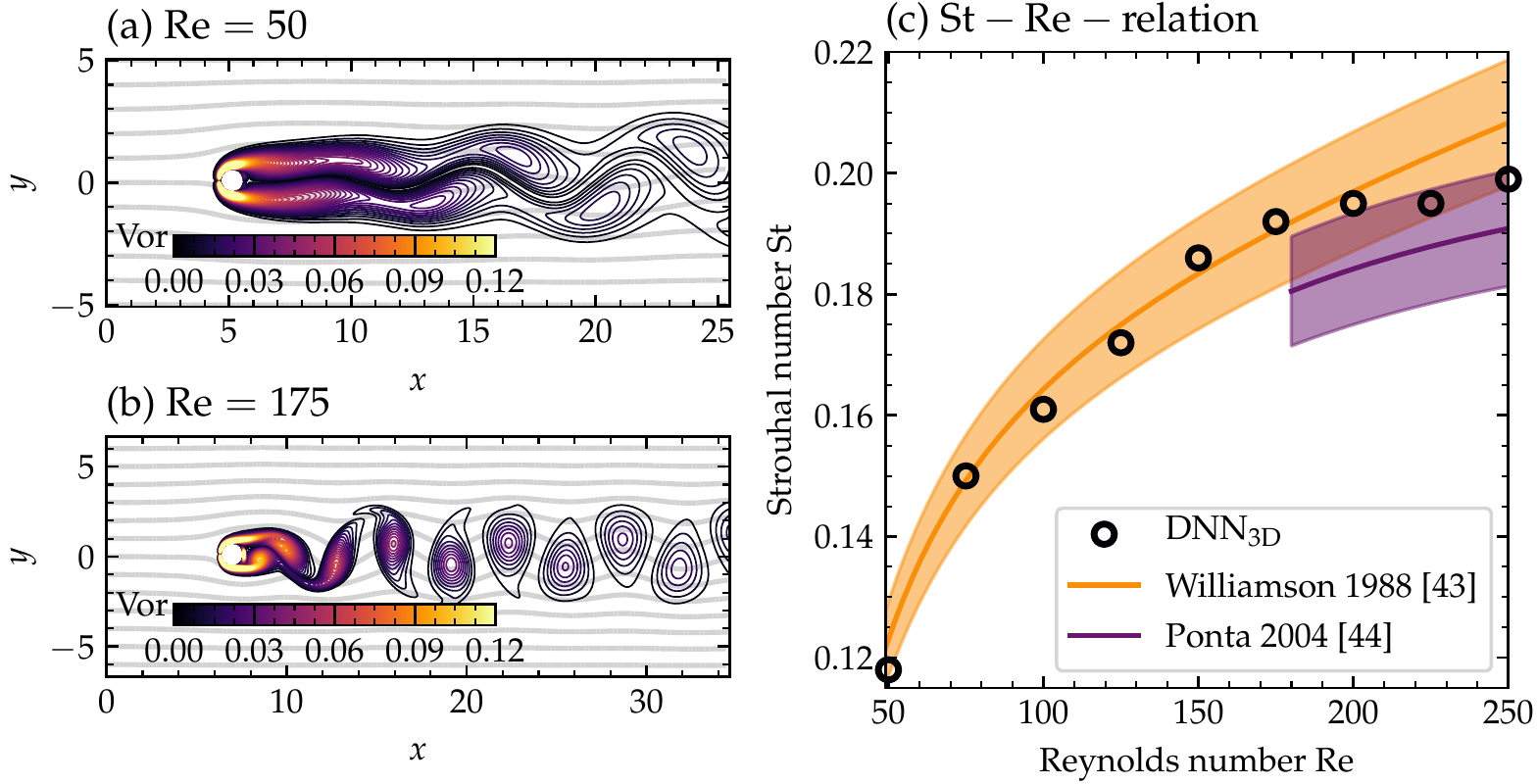}
\caption{(a) and (b) Contour plots of the absolute value of the vorticity in a flow with $\text{Re}=50$ and $\text{Re}=175$, respectively, past a cylindrical obstacle with the maximum value shown in white and the minimum value in black. The gray streamlines indicate that the domain is large enough for the boundary conditions to not come into play. The cross section was taken in the center of the three dimensional domain. (c) Strouhal versus Reynolds number in comparison with universal laws from the literature, i.e. \cite{williamson1988b} for $49 < \text{Re} < 180$ and \cite{ponta2004} for $\text{Re} \geq 180$. The shaded area indicates a 5\% uncertainty margin.}
\label{fig:karman_01}
\end{centering}
\end{figure*}

First we validate the implementation of the Navier-Stokes equation for an unsteady flow.
To do so, we simulate a von Kármán vortex street instability past a cylindrical obstacle with different Reynolds numbers, which was extensively studied theoretically, numerically, and experimentally \cite{vonkarman2004, williamson1988a, williamson1988b, williamson1989, williamson1996, ponta2004}. An inflow velocity of $u_i = 1$ is imposed at $x = 0$ in a cuboidal domain with dimensions $L_x\times L_y\times L_z$ with $25.5\leq L_x\leq34.7$, $10.1\leq L_y\leq3.4$, and $0.21\leq L_z\leq0.56$, using a discrete grid element size $\Delta x=0.04$ (for $\textrm{Re}\leq150$), and 0.035 (for $\textrm{Re}>150$). A cylindrical obstacle of axis $z$ and diameter $d = 1.75$ is set at $x \approx 9$. The boundaries parallel to the flow are set with free slip conditions, while the outflow at the upper $x$ boundary is free. The resulting oscillatory flow can be characterized by the Strouhal number $\text{St} = fd/u_i$, that uniquely depends on the Reynolds number, with $f$ being the outflow frequency. For $49 < \text{Re} < 180$ the Strouhal number follows a universal law \cite{williamson1988b}, while for higher Reynolds numbers the flow becomes three dimensional and follows a different law \cite{ponta2004}. Fig. \ref{fig:karman_01}a-b shows a snapshot slice through the three dimensional domain, illustrating iso-values of the vorticity magnitude for a Reynold number of 50 (a) and 175 (b). In Fig. \ref{fig:karman_01}c the predictions for St(Re) compare well with the universal laws assessed experimentally, with a deviation by less than 3\%.

\subsection{Steady state growth of an isolated dendrite}
\label{subsec:ivantsov}

Next, we test the growth of a free dendrite compared to the analytical Ivantsov solution \cite{ivantsov1947}, which relates the Pecl\'et number $\text{Pe} = RV/(2D)$ to the solute supersaturation as
\begin{align}
\label{eq:ivantsov_2D}
 \Omega &= \sqrt{\pi\text{Pe}}\,\exp{\left(\text{Pe}\right)}\,\text{erfc}\left(\sqrt{\text{Pe}}\right) =: \text{Iv}_{2\text{D}}(\text{Pe}) \nonumber\\
 &\text{with}\quad \text{erfc}\left(u\right) = 1 - \frac{2}{\sqrt{\pi}}\int_u^\infty\exp{(-s^2)}\,ds,\\
\label{eq:ivantsov_3D}
 \Omega &= \text{Pe}\,\exp{(\text{Pe})}\,\text{E}_1(\text{Pe}) =: \text{Iv}_{3\text{D}}(\text{Pe}) \nonumber\\
 &\text{with}\quad\text{E}_1(u) = \int_u^\infty\frac{\exp{(-s)}}{s}\,ds,
\end{align}
in 2D and 3D, respectively.

We simulate the growth of a single dendrite at a given imposed supersaturation until a steady state is reached, and compare the resulting P\'eclet number with the theoretical solution. 
We use a grid size of $230\times230\times230$ with a grid spacing $\Delta x$ that yields $\Delta x = l_D/15$ over a time of at least $50D/V^2_s$.
A unique needle is set to grow in the $x+$ direction and no-flux conditions are applied on all boundaries.
The needle is located either at the center (for $\Omega\leq0.02$) or at the edge (for $\Omega>0.2$) of the domain in $y$ and $z$.
The domain is shifted progressively to keep the location of the tip at $1/3$ of the domain length in $x$. 
The FIF integration radius is $r_\text{FIF}=5\Delta x$.

\begin{figure}[t]
\begin{centering}
\includegraphics[width=2.75in]{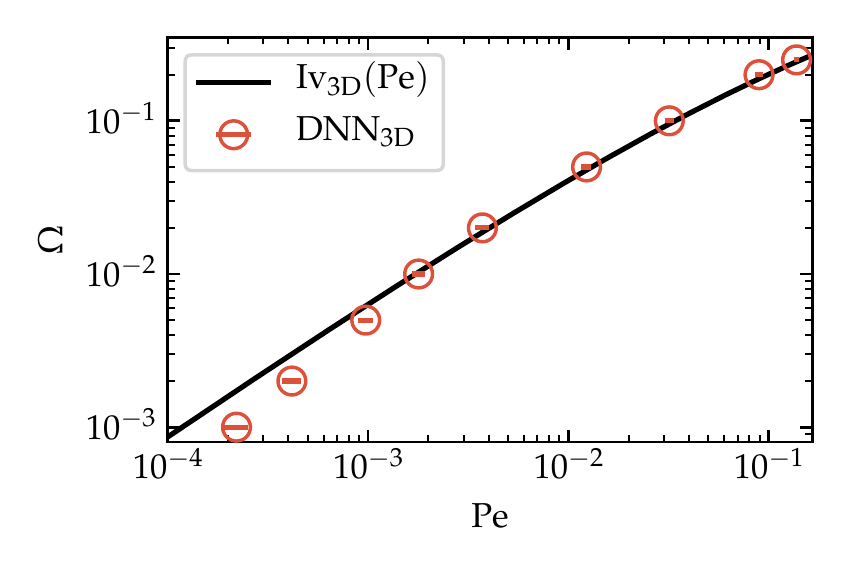}
\caption{Imposed supersaturation $\Omega$ versus P\'eclet number $\text{Pe}$ of a single isolated dendrite predicted by $3$D DNN simulations in comparison with the analytical Ivantsov law $\text{Iv}_{3\text{D}}(\text{Pe})$ \cite{ivantsov1947}. The grid spacing  $\Delta x$ was chosen so that the diffusion length $l_D = 15\Delta x$. Each value is an average over a time range that covers several oscillations. The error bars correspond to the standard deviation. \label{fig:ivantsov}}
\end{centering}
\end{figure}

Fig. \ref{fig:ivantsov} illustrates the comparison with the analytic law, showing that $\text{Pe}$, while slightly overestimated at low $\Omega$, the model essentially reproduces the expected steady growth velocity across orders of magnitude in the P\'eclet number.

%%%%%%%%%%%%%%%%%%%%%%%%%%%%%%%%%%%%%%%%%%%%%%%%%
\section{Equiaxed crystal growth in a forced flow}
\subsection{Simulations}

\begin{figure*}[!h]
\begin{centering}
\includegraphics[width=.7\textwidth]{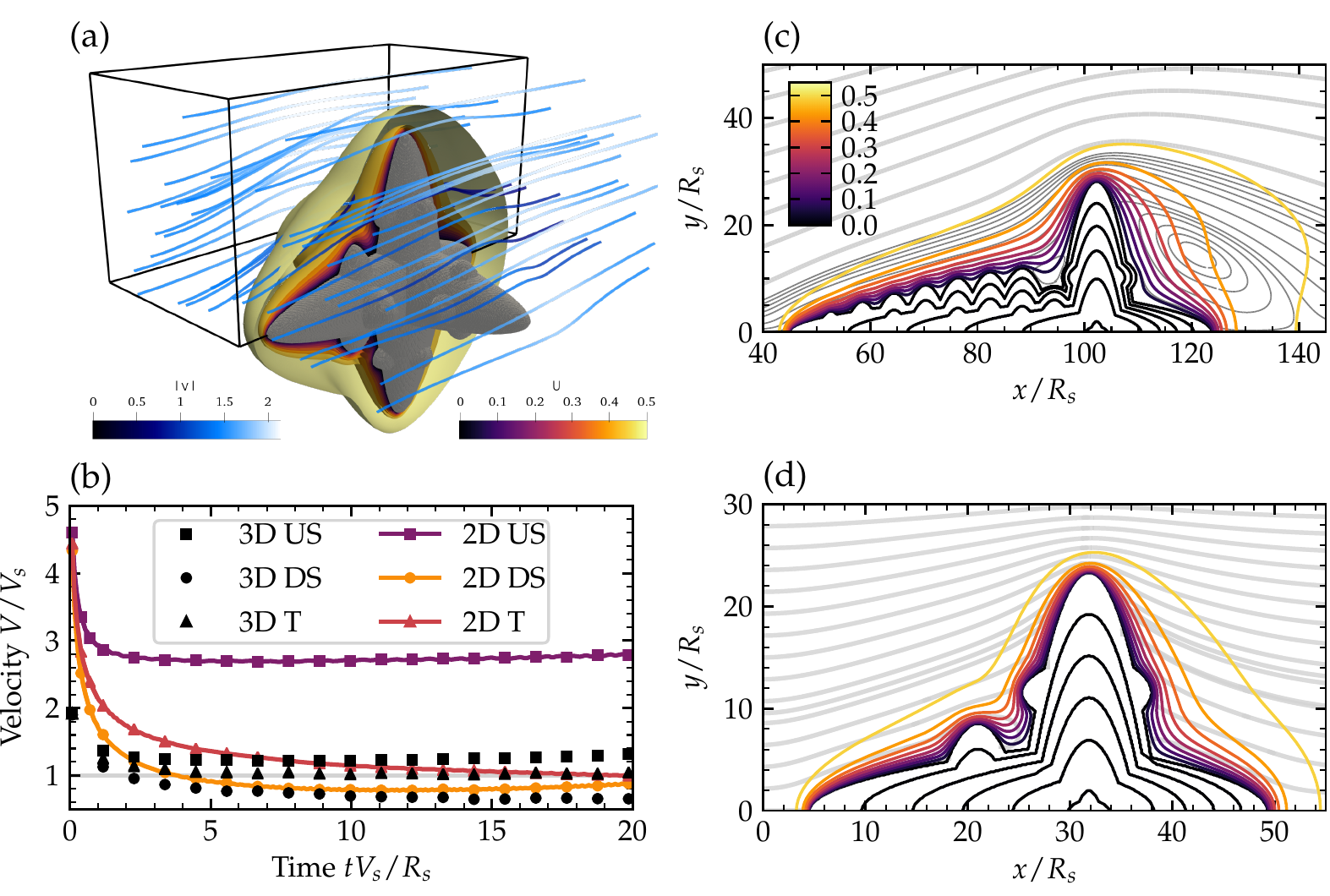}
\caption{(a) Dendritic grain (gray) under forced flow predicted by the DNN model at a time $20 R_s/V_s$. The iso-surfaces indicate the dimensionless solute field field $U$ from $0$ to $0.5$ with steps of $0.1$ and the streamlines show the laminar flow around the crystal. (b) and (d) show the cross section along the fluid flow direction. The colored contour lines indicate the field $U$; the black lines show the location of the solid-liquid interface at $t V_s/R_s = 0$, 4, 8, 12, 16, and 20. The gray streamlines represent iso-values of the stream function. The plots do not show the complete simulation domain. Plot (b) shows a higher density of thinner streamlines close to the dendrite in order to highlight the convective vortices next to the downstream tip. (c) Velocities of the upstream, downstream and transverse tips over time. The values correspond to average values over a small time range in order to smoothen the curves.
}
\label{fig:forced_flow}
\end{centering}
\end{figure*}

Finally, having independently verified the implementation of the Navier-Stokes equations and dendritic growth in the diffusive regime, we simulate the growth of equiaxed dendrites under forced flow. 
We perform simulations in both 2D and 3D for conditions close to those studied in 2D in \cite{tourret2019}.
We consider a model alloy of Schmidt number $\text{Sc}=\nu/D=20$, with $\nu=\eta/\rho$ the kinematic viscosity, at a solute supersaturation $\Omega = 0.5$.
We use a tip selection parameter $\sigma = 0.06$ in 2D and $\sigma= 0.04$ in 3D, which both correspond to a solid-liquid interface energy anisotropy of amplitude $\approx$0.01 for a one-sided model \cite{barbieri1989}.

For the $3$D simulation we use a domain with a size of $L_x\times L_y\times L_z = 51.0 R_s\times 25.4 R_s\times 25.4 R_s$ with a grid spacing of $\Delta x = 0.1 R_s$. 
An inflow with velocity $(u_i,0,0)$ is imposed at the left boundary ($x=0$), while the right boundary is set to allow free outflow, and the remaining boundaries hold free-slip conditions for the velocity (i.e. $v = \partial v/\partial y = 0$ in for $y$-boundaries and $w = \partial w/\partial z = 0$ for $z$-boundaries). Mirror symmetry conditions are applied for the diffusive field $U$ at all boundaries. 
Exploiting the symmetry of the domain and the expected laminar flow, we initiate a single equiaxed grain along the the $(y=z=0)$ edge of the domain, centered at $(25.5 R_s,0,0)$.
The grain consists four branches, with directions $x+$, $x-$, $y+$, and $z+$, and initial radii and lengths of $0.4 R_s$ and $2 R_s$, respectively. 
For the 2D simulation, we use similar boundary conditions and grid spacing but with a size of $L_x\times L_y = 205 R_s\times 102 R_s$. The grain is generated at $(102 R_s,0)$. 

For similar physical parameters, the expected steady state velocities $V_s$ in 2D and 3D differ, such that the scaled inflow velocities $u_i/V_s$ also differ.
Combining microscopic solvability \eqref{eq:microscopic_solvability} with the definition of the P\'eclet number, $\text{Pe} = RV/(2D)$, one can write $V_sd_0/D = 2\sigma P_s^2$.
For $\Omega=0.5$, since the P\'eclet number is {0.1873} in 2D and {0.6101} in 3D, the steady growth velocity $V_s$ in 3D is about 7 times higher than in 2D. 
The inflow velocity is set to $u_i=0.0421/d_0$, which corresponds to $u_i/V_s = 10.0$ in 2D and $u_i/V_s=1.41$ in 3D.

The total simulation time is chosen as $20R_s/V_s$, which was found to be sufficient to achieve steady-state growth.
For both $2$D and $3$D simulations, the numerical parameters are $K_{\Delta t} = 0.6$, $\omega_\text{up} = 0.9$, $\omega_\text{SOR} = 1.7$ and $\overline{r}_\text{SOR} = 10^{-3}$ (see Ref. \cite{tourret2019} for details). The FIF integration domain has a radius $r_\text{FIF} = 5\Delta x$. Side branches are generated every time a needle grows by $10 R_s$.

\subsection{Results}

The 3D simulation was performed in under 20~h with a single Nvidia RTX 2080TI GPU, and the results are illustrated in Fig. \ref{fig:forced_flow}.
Fig. \ref{fig:forced_flow}a shows the grain shape in the 3D simulation after a time $20R_s/V_s$, with iso-surface of the solute field and streamlines of the flow.
Fig. \ref{fig:forced_flow}{b} shows the evolution of the upstream (US), downstream (DS), and transverse (T) tips in both 2D and 3D simulations.
Fig. \ref{fig:forced_flow}{c-d} shows the interface, solute field, and streamlines in 2D {(c)} and along the $(z=0)$ plane in 3D (d).

Qualitatively, the effect of fluid flow on the tip velocities is similar, namely the upstream tip velocity is increased, the downstream tip velocity is decreased, and the steady transverse tip velocity is barely affected.
However, the amount of change differs substantially between 2D and 3D simulations.
The upstream velocity increase is much higher in 2D (about 3 times higher than the transverse tip velocity) than in 3D (increased by about 40\%).
This is not only due to the the different scaling of $V_s$ in 2D and 3D, but also to the resulting flow pattern.
As already pointed out in previous studies \cite{jeong2001,jeong2003,lu2005,sakane2018}, in 3D the flow can easily pass by the transverse arm while in 2D the arm acts as a wall that the flow has to entirely go around.
As a result, the flow velocities are overestimated in 2D, and flow patterns are also importantly affected. 

An important consequence on the flow pattern is the formation of convective vortices around the downstream arm, which only appear in 2D.
An outcome of these 2D vortices is that the downstream tip accelerates as the vortices feed it solute, and hence the tip cannot reach a steady state.
Meanwhile, in 3D, the downstream tip seems to reach a well-defined steady growth velocity. 

%%%%%%%%%%%%%%%%%%%%%%%%%%%%%%%%%%%%%%%%%%%%%%%%%
\section{Summary and outlook}
\label{sec:summary}

We presented the first three-dimensional implementation of the Dendritic Needle Network (DNN) model for binary alloy isothermal solidification with fluid flow. The code implementation was validated for unsteady oscillatory flow past an obstacle, and verified for steady state growth in the diffusive regime. Then, we performed simulations of equiaxed growth of a single grain in a forced flow and compared results of 2D and 3D simulations. The results show that the growth dynamics significantly deviates from diffusive solidification. Furthermore, the acceleration of the upstream tip and deceleration of the downstream tip, differ significantly in the 2D and 3D cases, in agreement with previous studies \cite{jeong2001,jeong2003}. These results further highlight the importance of 3D simulations in order to produce results that can be compared to experiments or solidification processes on a quantitative basis.

In the future, we expected the DNN model to provide computationally efficient and spatially extended simulations of solidification in a low P\'eclet number regime that remains challenging to established simulation methods such as phase-field.
This should allow exploring the mechanisms of microstructure selection at the scale of thousands of dendrites \cite{tourret2015jom, tourret2016}.
Ongoing and upcoming developments from this work include: quantitative comparison to phase field simulations \cite{jeong2001,jeong2003}, study of the effect of the relative orientations of crystal and flow \cite{sakane2018,badillo2007}, and the extension to directional solidification conditions \cite{tourret2013a,tourret2016} in order to study, for instance, the selection of dendritic spacings in the presence of micro- or hyper-gravity conditions \cite{steinbach2009,viardin2019}, or the selection of grain boundary orientations during columnar grain growth competition \cite{tourret2015growth, tourret2017grain}.

%%%%%%%%%%%%%%%%%%%%%%%%%%%%%%%%%%%%%%%%%%%%%%%%%%%%%%%
\section*{Acknowledgements}
This work was supported by the European Union’s Horizon 2020 research and innovation programme through DT’s Marie Skłodowska-Curie Individual Fellowship (Grant Agreement 842795).

%\section*{References}

\bibliographystyle{unsrt}
\bibliography{2020_Isensee_MCWASP}

\end{document}